\documentclass[8pt]{statsoc}
\usepackage[left=2cm,right=2cm,top=2cm, bottom=2cm]{geometry}
\usepackage{color}
\usepackage{apalike}
\usepackage{verbatim}
\usepackage{framed}
\usepackage{wrapfig}
\usepackage{caption}
\usepackage{floatrow}
\usepackage{graphics}
\usepackage{amsmath}
\usepackage{setspace}
\usepackage{graphicx}
\def\ci{\perp\!\!\!\perp}

\title[Weighing evidence with the Meta-Analyzer]{On the physical interpretation of a meta-analysis in the presence of heterogeneity and bias: from clinical trials to Mendelian randomization}

\author[J Bowden {\it et. al.}]{Jack Bowden$^{1,2*}$ }
\address{$^{1}$ MRC Integrative Epidemiology Unit, University of Bristol, UK.\\
$^{2}$ MRC Biostatistics Unit, Cambridge.}
\email{jack.bowden@mrc-bsu.cam.ac.uk}
\author[J Bowden {\it et. al.}]{Chris Jackson}
\address{MRC Biostatistics Unit, Cambridge, UK}
\begin{document}

\maketitle

\begin{abstract}
{The funnel plot is a graphical visualisation of summary data estimates from a meta-analysis, and is a useful tool for detecting departures from the standard modelling assumptions. Although perhaps not widely appreciated, a simple extension of the funnel plot can help to facilitate an intuitive interpretation of the mathematics underlying a meta-analysis at a more fundamental level, by equating it to determining the centre of mass of a physical system. We used this analogy, with some success, to explain the concepts of weighing evidence and of biased evidence to a young audience at the Cambridge Science Festival, without recourse to precise definitions or statistical formulae. In this paper we aim to formalise this analogy at a more technical level using the estimating equation framework: firstly, to help elucidate some of the basic statistical models employed in a meta-analysis and secondly, to forge new connections between bias adjustment in the evidence synthesis and causal inference literatures.}
\end{abstract}

\section{Introduction}

Let $y_{i}$, $i=1,...,k$, represent summary estimates of the same apparent quantity from $k$ independent information sources in a meta-analysis. The $i$'th estimate is associated with a fixed and known variance, $s^{2}_{i}$. The standard fixed effect model assumes
\begin{equation}
y_{i} =  \mu + s_{i}\epsilon_{i},\quad \epsilon_{i}\sim N(0,1)
\label{eq:REadd}
\end{equation}
The focus for inference, in terms of point estimates and confidence intervals is the population mean effect parameter $\mu$. The fixed effect estimate $\hat{\mu}$ and its variance are given by the well known formula
\begin{equation}
\hat{\mu} = \frac{\sum^{k}_{i=1} w_{i}y_{i}}{\sum^{k}_{i=1} w_{i}}, \quad \text{Var}(\hat{\mu}) = 1/\sum^{k}_{i=1} w_{i}.
\label{eq:FEest}
\end{equation}
Meta-analysis has a long history in medical research, where the information source has traditionally been a randomized clinical trial, comparing an experimental treatment against standard therapy. However, its use spans the entire scientific spectrum, for example in the areas of psychology (Hedges, 1992); inter-laboratory experiments (Paule and Mandel, 1982); particle physics (Baker and Jackson, 2013);  Mendelian randomization (Burgess and Thompson, 2010) and many others.\\
\\
Regardless of the application area, if $s_{i}$ and $\epsilon_{i}$ are mutually independent, then equation (\ref{eq:FEest}) will give a consistent estimate for  $\mu$. In order to provide a convenient shorthand that easily generalizes to more complex models, we write $\ci_{(s_{i},\epsilon_{i})}$ to denote this assumption.

\subsection{The funnel plot and the fixed effect model}

The funnel plot (Sterne and Egger, 2001) is a graphical visualisation of data from a meta-analysis. It plots study $i$'s effect size on the x-axis versus a measure of precision on the y-axis (usually 1/$s_{i}$). The independence assumption  $\ci_{(s_{i},\epsilon_{i})}$ of model (\ref{eq:REadd}) implies that there should be no correlation between effect size and precision. If this is the case, then the plot should appear symmetrical, in that the results from smaller studies should funnel in towards those from larger studies. It therefore provides a simple means to visually assess the plausibility of $\ci_{(s_{i},\epsilon_{i})}$.\\
\\
Although perhaps not as widely appreciated, a simple extension of the funnel plot can help to facilitate an intuitive interpretation of the mathematics underlying a meta-analysis at a more fundamental level, by equating it to determining the centre of mass of a physical system.  This analogy was recently exploited by MRC scientists to build a machine (named the `Meta-Analyzer') to explain the concepts of weighing evidence and of biased evidence to a young audience at the Cambridge Science Festival, without recourse to precise definitions or statistical formulae. We have since produced a web-emulation of the machine to bring these ideas to an even wider audience of students and researchers - it can be found at \verb https://chjackson.shinyapps.io/MetaAnalyser/ .

\begin{figure}[hbtp]
 \centering
\includegraphics[width=0.6\textwidth,clip]{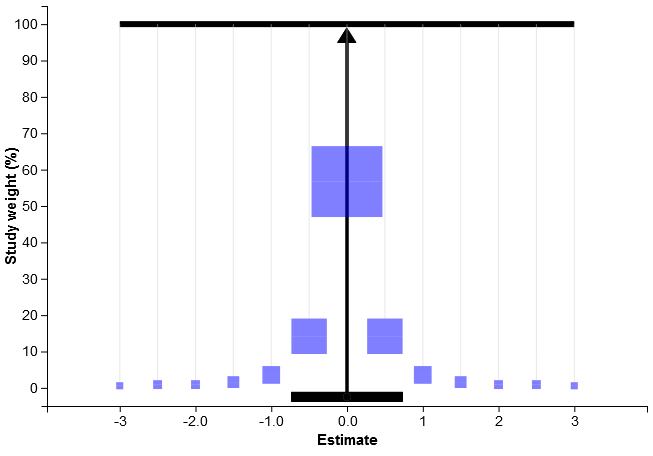}
\caption{{\it The meta-analyzer supporting a fictional body of 13 study results. The centre of mass (overall estimate) is located at zero.}}
\label{fig:symmetric}
\end{figure}

Figure \ref{fig:symmetric} shows the Meta-Analyzer web app populated with a fictional meta-analysis of $k$=13 studies. In it the standard funnel plot has been augmented so that the  area representing point $i$ is proportional to study $i$'s fixed effect weight $w_{i}$ = 1/$s^{2}_{i}$, in order to promote its interpretation as a physical mass. It is proportional because, for study $i$, we show its weight in the Meta-Analyzer as a percentage of the total weight in the analysis, which equals 100$\times$($w_{i}/\sum^{k}_{i=1}w_{i}$).\\
\\
Points are joined by horizontal cord to a pole that is joined itself to a vertical stand at a pivot point, $p$ say. Study weights are imagined to exert a downward force due to gravity. Since the pole is perfectly horizontal, it is intuitively understood that $p$ satisfies the {\it physical law}:
\[\
\sum^{k}_{i=1}w_{i}(y_{i} - p) = 0,
\]
and is therefore equal to the centre of mass (Beatty, 2005). The above formula can be viewed as a rudimentary estimating equation, a construction we continue to utilize throughout this paper. It is simple to verify that $p$ is identical to the fixed effect estimate $\hat{\mu}$ in (\ref{eq:FEest}). The length of the stand base along the x-axis shows the 95\% confidence interval for $\mu$. Again, there is a nice physical analogy to draw. When there is a lot of uncertainty as to the overall effect estimate, so that Var($\hat{\mu}$) is large, the stand length must be wide in order to properly account for its instability. Conversely, when the overall effect estimate is very precise, so that Var($\hat{\mu}$) is small, the stand length need only be short. When populated with such idealised data as in Figure \ref{fig:symmetric}, the Meta-Analyzer may remind some of Galton's bean board or quincunx, a tool that also exploits gravity to illustrate how statistical laws can emerge from a seemingly random physical process. \\
\\
Although the Meta-Analyzer was intended to be a simple educational tool to demonstrate the basics of evidence synthesis, we have continued to find the physical system it describes useful more broadly, to easily explain some common secondary issues in meta-analysis and to help make new connections between statistical techniques for bias adjustment from the literature that seem, at first sight, unrelated. In Section 2 we show that it helps to transparently demonstrate the implications of moving from a fixed to a random effects model and to assess the influence of outlying studies. In Section 3 we discuss  the issue of biased evidence, how the Meta-Analyzer was used to explain this concept to a lay audience, and how small study bias is commonly addressed by medical statisticians using Egger regression (Egger et al, 1997). In Section 4, we review the method of Mendelian randomization -- a technique for estimating the causal effect of a modifiable exposure on a health outcome using observational data, by circumventing the problem of confounding. We draw parallels between adjusting for bias in Mendelian randomization and small study bias in meta-analysis, and describe an extension to the standard model assumed by Egger regression first proposed in Bowden, Davey Smith and Burgess (2015) within the context of Mendelian randomization, that can (in theory) be applied to both fields. In Section 5 we show that, by viewing Egger regression from a causal inference perspective, a novel estimating equation interpretation of this method is found that can be intuitively visualized via the Meta-Analyzer. We illustrate this new interpretation on some real data examples and conclude with a discussion in Section 6.

\section{Random effects models}

A common issue in meta-analysis is accounting for between study heterogeneity. In a fixed effect meta-analysis all studies are assumed to provide an estimate of the same quantity, and the only difference between studies is in the precision of their respective estimates for this quantity. This is often thought to be an over-simplistic model, especially when flatly contradicted by Cochran's Q-statistic, $Q$ = $\sum^{k}_{i=1}\frac{1}{s^{2}_{i}}(y_{i} - \hat{\mu})^{2}$, being substantially larger than its expected value of $k$-1 under model (1), and a random effects model is preferred. Two distinct approaches for incorporating heterogeneity have emerged. The first, and most popular is via an additional additive random effect, as in model (\ref{eq:RE}) below (e.g. DerSimonian and Laird, (1986), Higgins and Thompson, (2002)). The second is via the addition of a multiplicative scale factor, as in model (\ref{eq:MRE}) below (for example Thompson and Sharp (1999), Baker and Jackson (2013)).

\begin{eqnarray}
y_{i} &=&  \mu + s_{i}\epsilon_{i} + \delta_{i}, \quad \delta_{i} \sim N(0,\tau^{2}), \quad \epsilon_{i}\sim N(0,1). \label{eq:RE} \\
y_{i} &=&  \mu + \phi^{\frac{1}{2}}s_{i}\epsilon_{i}, \quad \epsilon_{i}\sim N(0,1). \label{eq:MRE}
\end{eqnarray}

We first note that, regardless of whether model (\ref{eq:RE}) or (\ref{eq:MRE}) holds in practice, under the independence assumption $\ci_{(\delta_{i},s_{i},\epsilon_{i})}$,  consistent estimation of the overall mean parameter $\mu$ is achieved by fitting the fixed effect model (\ref{eq:REadd}). In practice, estimation of $\mu$ can follow by simple application of formula (\ref{eq:FEest}) except now the weight given to study $i$, $w_{i}$, changes to $\frac{1}{s^{2}_{i}+\tau^{2}}$ and $1/\phi s^{2}_{i}$ under models (\ref{eq:RE}) and (\ref{eq:MRE}) respectively.  The point estimate for $\mu$ obtained by fitting model (\ref{eq:MRE}) is identical to that obtained from fitting model (\ref{eq:REadd}). This is because the common term $\phi$ simply cancels from the numerator and denominator in (\ref{eq:FEest}) and only the variance of the estimate is altered. However, both the point estimate and variance for $\hat{\mu}$ change under the additive random effect model (\ref{eq:RE}). Because its use is ubiquitous, especially in the field of medical research (Baker and Jackson, 2013), we focus only on the additive model for the remainder of this section.\\
\\
If, after fitting the fixed effect model (\ref{eq:REadd}), Cochran's $Q$ statistic about $\hat{\mu}$ indicates additional heterogeneity ($Q$ $>$ $k-1$), then it is common practice to estimate $\tau^{2}$ using the procedure defined by DerSimonian and Laird (1986), to give $\hat{\tau}^{2}_{DL}$. This estimate conveniently provides a link between $Q$ and a popular measure of heterogeneity, $I^2$, (Higgins, 2002) as follows:
\begin{equation}
\frac{Q - (k-1)}{Q} = \frac{\hat{\tau}^{2}_{DL}}{\hat{\tau}^{2}_{DL}+s^{2}_{typ}} =  I^{2}, \nonumber
\end{equation}

where $s^{2}_{typ}$ is referred to as the `typical' within study variance.

\subsection{Random effects models via the Meta-Analyzer}

In order to provide a physical interpretation of the calculations underpinning a meta-analysis so that they can be implemented using the Meta-Analyser, we formulate a system of estimating equations (as a natural progression of the single estimating equation for fixed effect model (1)) to fit the random effects model to find $\mu$ and $\tau^{2}$ as below:

\begin{eqnarray}
\text{Weight equation:} \quad w_{i} &=& 1/(s^{2}_{i} + \tau^{2}) \label{eq:PMw} \\
\text{Mean equation:} \quad \sum^{k}_{i=1}w_{i}(y_{i} - \mu) &=& 0 \label{eq:PMmu} \\
\text{Heterogeneity equation:} \quad \sum^{k}_{i=1}w_{i}(y_{i}-\mu)^{2} - (k-1) &=& 0.\label{eq:PM}
\end{eqnarray}

\begin{figure}[hbtp]
 \centering
\includegraphics[width=0.49\textwidth,clip]{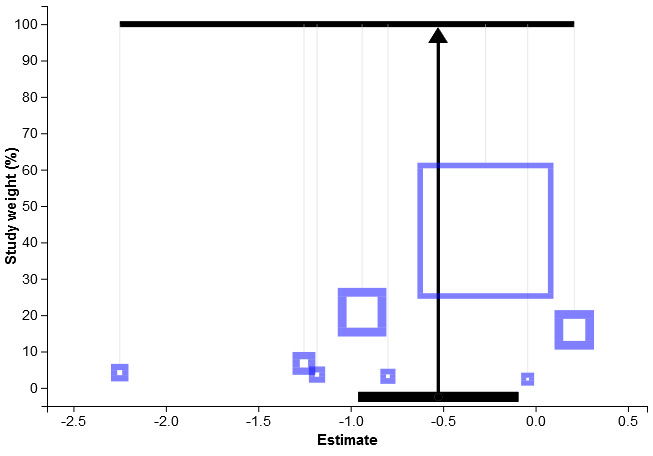}
\includegraphics[width=0.49\textwidth,clip]{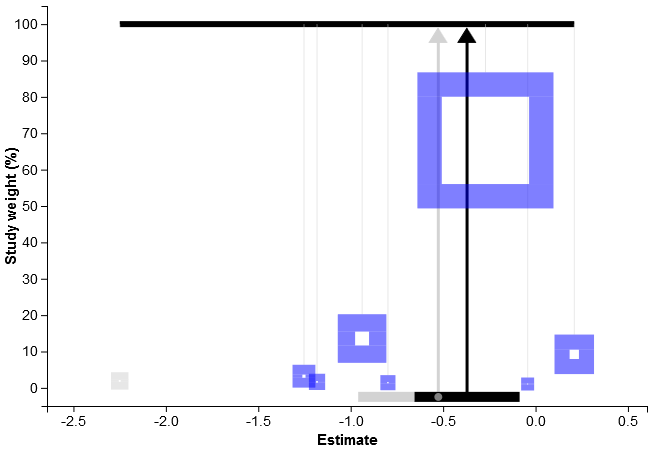}
\caption{{\it The Meta-Analyzer supporting the magnesium data under a random effects model for all trials (left); and with the Shechter trial removed (right).}}
\label{fig:MetaShiny}
\end{figure}

\begin{figure}[hbtp]

\begin{center}
\includegraphics[bb=0 228 630 574,width=0.4\textwidth,clip]{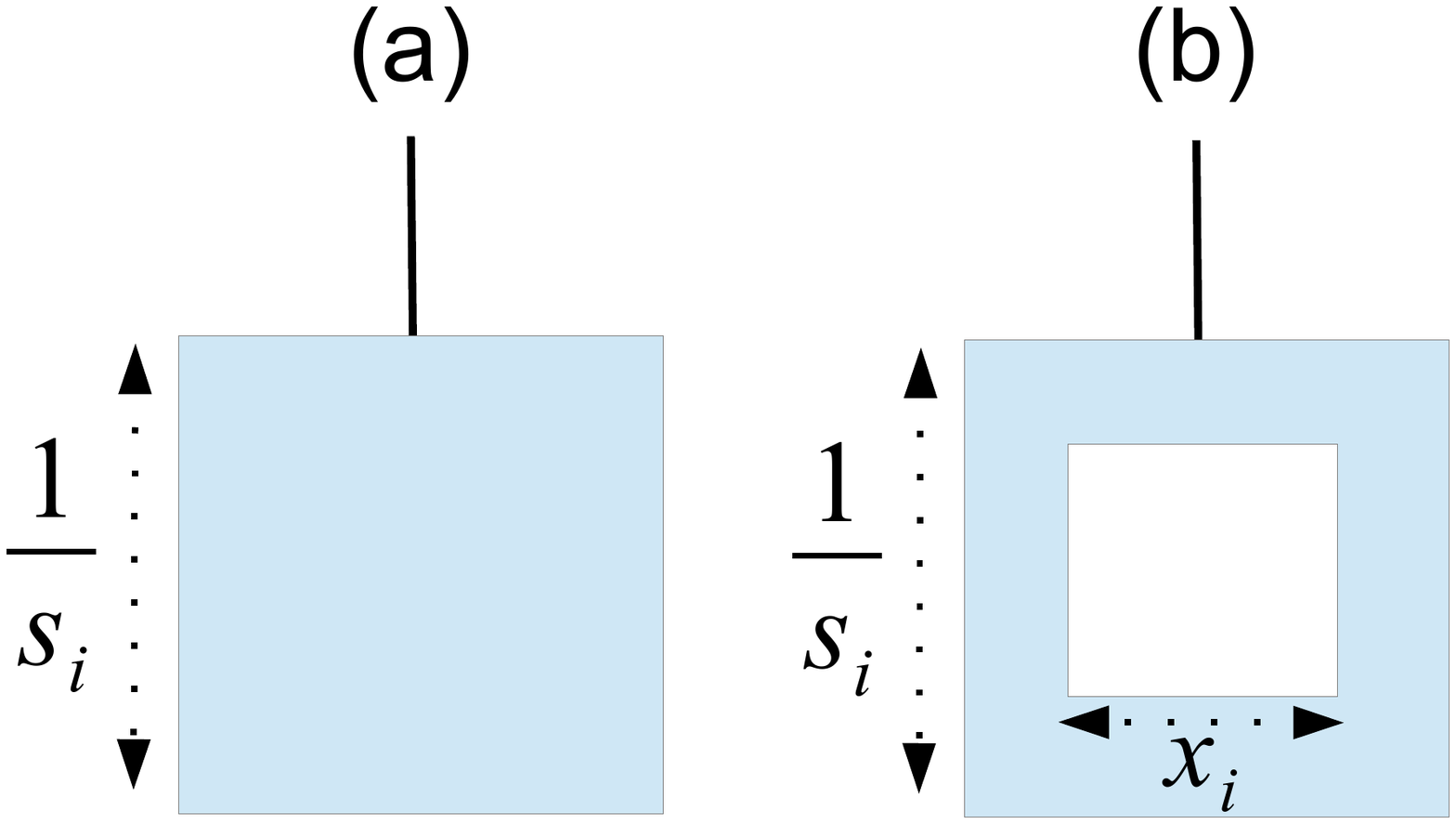}
\captionof{figure}{\it (a) Weight given to study $i$ in a fixed effect meta-analysis. (b) Weight given to study $i$ in an additive random effects meta-analysis. A Square of length $x_{i}$ is removed from weight $i$ in order to satisfy the pythagorean formula.}
\label{fig:Drill}
\end{center}
\end{figure}

Formula (\ref{eq:PM}) is referred to as the generalized $Q$ statistic (Bowden et al, 2011) and, when solved in conjunction with (\ref{eq:PMw}) and (\ref{eq:PMmu}), it returns an estimate for $\mu$ and the Paule-Mandel (PM) estimate for $\tau^{2}$ (Paule and Mandel, 1982), which we denote by $\hat{\tau}_{PM}$. As with $\hat{\tau}^{2}_{DL}$ the PM estimate is constrained to be positive, it is known to provide a more reliable estimate for the between study heterogeneity than $\hat{\tau}^{2}_{DL}$ (Veroniki et al, 2015). Random effects model (\ref{eq:RE}) has been promoted by the Cochrane collaboration (Higgins and Green, 2011) and formally justified as a basis for inference beyond the current meta-analysis to future studies and populations (Higgins, Thompson and Speigelhalter, 2009). It reduces to the fixed effects model when $\tau^{2}$ is either fixed or estimated to be 0. This implies $\mu_{i} \equiv \mu$ for all $i$.\\
\\
Application of the random effects model, with additional variance component $\tau^{2}$, leads to study results being both down-weighted and more similarly weighted. Furthermore, the original weight given to large studies is reduced to a greater extent than those of smaller studies. This issue is fairly subtle and hard to comprehend, but the Meta-Analyzer provides a very simple visualization, by adjusting the mass of each weight. Figure \ref{fig:MetaShiny} (left) shows the Meta-Analyzer populated with a data set of 8 randomized trial results, each assessing the use of magnesium to treat myocardial infarction. The effect measure, $y_{i}$, is the log-odds ratio of death between treatment and control groups for study $i$. These data were previously analyzed by Higgins and Spiegelhalter (2002b).\\
\\
Between trial heterogeneity is present for these data ($\hat{\tau}^{2}_{DL}$ = 0.095, $I^{2}$ = 27.6\%). Under random effects model (3) $w_{i}$ is reduced from $\frac{1}{s^{2}_{i}}$ to 1/($s^{2}_{i}$ + $\hat{\tau}^{2}$). We represent the weight `loss' induced by moving from a fixed to an additive random effects model by `drilling out' a square of length $x_{i}$, in order to satisfy the pythagorean identity $x^{2}_{i}+ \frac{1}{s^{2}_{i}+\hat{\tau}^{2}}$ = $\frac{1}{s^{2}_{i}}$, as illustrated in Figure \ref{fig:Drill}. The centre of mass defined by the holed-out weights and calculated by the Meta-Analyzer is automatically consistent with the random effects estimate for $\mu$. In Figure \ref{fig:Drill} we plug in $\hat{\tau}^{2}_{DL}$ for $\hat{\tau}^{2}$ for illustration. Full results obtained via the estimating equation approach (and so using the PM estimate for $\tau^{2}$) are shown in Table \ref{tab:MagResults}.

The fact that large studies lose more of their relative weight than small studies under an additive random effects model is immediately apparent from Figure \ref{fig:MetaShiny} (left). We note briefly that no holing out is necessary when the multiplicative random effect model (\ref{eq:MRE}) is used. This is because the constant factor $\phi$ does not alter the weight given to study $i$ as a proportion of the total weight in the analysis, whatever its value. However, when $\phi \neq $ 1 then the variance of $\hat{\mu}$ will differ to that of the fixed effect model and so the stand length (confidence interval) will subsequently change.

\subsection{Sensitivity to outliers}

The amount of heterogeneity estimated in a meta-analysis can depend heavily on extreme, and often small, study results (Bowden et al, 2011). It is therefore useful in some circumstances to perform a sensitivity analysis, in which an outlying study result is excluded. Figure \ref{fig:MetaShiny} (right) shows the Meta-Analyzer supporting the magnesium data under random effects model (3) excluding the Shechter study (shown in grey). The solid black support stand in Figure \ref{fig:MetaShiny} (right) shows the overall estimate and corresponding 95\% confidence interval in this case. The grey support stand shows the original point estimate and confidence interval. In this example, exclusion of the outlying study removes a large proportion of the between trial heterogeneity (updated $\hat{\tau}^{2}_{DL}$ = 0.012, $I^{2}$ = 5.2\%), making interpretation of the remaining trial data easier. Our web application facilitates easy transitions between various models like this as part of a sensitivity analysis. Users also see the Meta-Analyzer dynamically tip and re-balance in response to their latest analysis choice.

\begin{table}[htbp]
\begin{center}
\begin{tabular}{lcccc}
\hline
Model & \\
Parameter & Est & S.E & t value & p-value \\
\hline
\multicolumn{5}{c}{All studies} \\
$\mu$  & -0.516 & 0.214 &-2.408 & 0.047 \\
$\tau^{2}_{PM}$ & 0.084 &-&-&-\\
$\tau^{2}_{DL}$ & 0.095 ($I^2$ = 27.6\%) &-&-&-\\
&&&&\\
\multicolumn{5}{c}{Shechter study removed} \\
$\mu$  & -0.362 &0.219& -1.653&  0.149 \\
$\tau^{2}_{PM}$ & 0.008 &-&-&-\\
$\tau^{2}_{DL}$ & 0.012 ($I^2$ = 5.1\%) &-&-&-\\
\hline
\end{tabular}
\end{center}
\caption{\label{tab:MagResults} {\it Meta-analysis of the Magnesium data under random effects model (3), with and without the Shechter trial.}}
\end{table}

\section{Small study bias}

\subsection{The Aspirin data}

\begin{figure}[hbtp]
 \centering
\includegraphics[width=0.7\textwidth,clip]{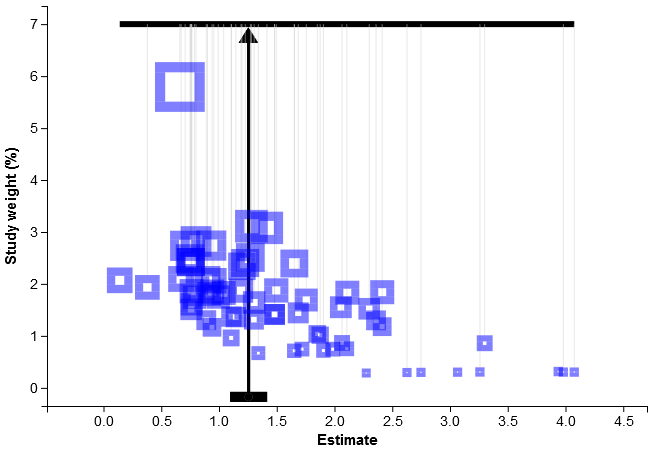}
\caption{{\it Meta-Analyzer supporting the Aspirin data under a random effects model.}}
\label{fig:ShinyFunnel2}
\end{figure}

Figure \ref{fig:ShinyFunnel2} (left) shows the Meta-Analyser enacted on 63 randomized controlled trials reported by Edwards et al. (1998) that each investigated the benefit of oral Aspirin for pain relief. Study estimates $y_{i}$ represent the log-odds ratios for the proportion of patients in each arm who had at least a 50\% reduction in pain. Between trial heterogeneity was present for these data ($\tau^{2}_{DL}$ = 0.04, $I^{2}$ = 10\%) and Figure \ref{fig:ShinyFunnel2} (left) reflects the weight given to each study by the Meta-Analyzer under the random effects model (\ref{eq:RE}) using $\tau^{2}_{DL}$ as the heterogeneity parameter estimate. Despite the apparent between trial heterogeneity, the conclusion of the random effects meta-analysis is that oral Aspirin is an effective treatment, the combined log-odds ratio estimate is 1.26 in favour of Aspirin with a 95\% confidence interval (1.1,1.41). Full results are shown in Table \ref{tab:AspirinResults}.

The hypothetical data shown in Figure 1 is perfectly symmetrical about its centre of mass, indicating that there is no correlation between effect size and precision across studies. However, there is a clear asymmetry present in the Aspirin data, smaller studies tend to show larger effect size estimates, whereas larger studies tend to report more modest results. For these data, Cor$(y_{i},1/s_{i})$ =  -0.7, which suggests $\ci_{(\delta_{i},s_{i},\epsilon_{i})}$ does not hold under model (\ref{eq:RE}). The phenomenon of observing a negative correlation between study precision and effect size is often given the umbrella term `small study bias' (Egger et al., 1997; Sterne et al., 2011; R$\ddot{u}$cker et al., 2011).

\subsection{The causes and consequences of small study bias}

Small study bias could actually be caused by real differences between small and large studies. Small trials may employ a more intensive intervention and therefore generate a greater effect on disease outcomes than larger trials (Egger et al, 1997; Bowater and Escrela, 2013). Asymmetry could also be a simple artefact of the data. For example, point estimates are not strictly independent of their estimated variances when calculated from binary or count outcomes, (Harbord et. al, 2006, Peters et. al, 2010). However, its cause could be also be more sinister. Publication bias, or the file-drawer problem (Rosenthal, 1979) occurs when journals selectively publish study results that achieve a high level of statistical significance, and also induces asymmetry.\\
\\
Much attention has been focused on methods to adjust for small study bias assuming that it is caused by selective dissemination of research findings, a practice which unfortunately is prevalent in biomedical research (Dwan et al, 2008). A common strategy is to propose an underlying selection model that lead to the generation of biased data. Many have followed Hedges (1984,1992), Copas (1999) and Copas and Shi (2000) in assuming the probability of observing a study result is some function of its precision. Whatever its true cause, when small study bias is present it can severely and adversely affect the conclusions reached. For example, it can lead one to detect apparent between trial heterogeneity when, in truth, none exists, and it can induce substantial bias into the overall estimate, $\hat{\mu}$, particularly under the random effects model (\ref{eq:RE}), because it gives more relative weight to small studies than the fixed effect model (Henmi and Copas, 2010, Bowden et al, 2011). Indeed, for the Aspirin data, we see a slight reduction in the log-odds ratio estimate under the multiplicative random effects model (Table \ref{tab:AspirinResults}). In the case of the Aspirin data, it is reasonable to suspect that the effect of oral Aspirin on pain relief is substantially smaller than suggested by either random effects analysis.

\subsection{Small study bias explained to a lay audience via the Meta-Analyzer}

When attempting to explain the concept of biased evidence and of bias adjustment to the science festival audience, we opted for a simplified version of Trim and Fill (Duval and Tweedie, 2000). It aims to replace `missing studies' in a meta-analysis by a process of reflection, until symmetry in the funnel plot is restored. We illustrated their idea within the context of a Sherlock Holmes' style mystery (see Box 1 and Figure 9 in the Appendix which shows the Meta-Analyzer at its initial conceptual state and in situ at the Cambridge Science Festival (\verb http://www.sciencefestival.cam.ac.uk/ ).

\begin{table}[htbp]
\begin{center}
\begin{tabular}{lcccc}
\hline
Model & \\
Parameter & Est & S.E & t value & p-value \\
\hline
\multicolumn{5}{c}{Random effects model (3)} \\
$\mu$ & 1.26 &0.082 &15.4 &$<$10e-16 \\
$\tau^{2}$& 0.04 &-&-&-\\
\multicolumn{5}{c}{Random effects model (4)} \\
$\mu$ &1.23 &0.080 &15.4 & $<$10e-16 \\
$\phi$& 1.11  &-&-&-\\
\multicolumn{5}{c}{Egger regression model (10)} \\ \\
$\beta_{0}$ & 2.11   & 0.31 & 6.77 & 5.8e-09\\
$\mu$       & 0.025 & 0.19 & 0.13 & 0.89\\
$\phi$      & 0.64 &-&-&-\\
\hline
\end{tabular}
\end{center}
\caption{\label{tab:AspirinResults} {\it Results for Meta-analyses of the Aspirin data.}}
\end{table}

\subsection{Egger regression}

Despite the intuition and appeal of its end result, the mathematics behind Trim and Fill are quite complicated. Perhaps due to its relative simplicity, the most popular approach to testing and adjusting for small study bias in medical research is Egger regression (Egger et al., 1997). This assumes the following linear fixed effect model in order to explain the correlation between $y_{i}$ and $1/s_{i}$:
\begin{equation}
\frac{y_{i}}{s_{i}} =\beta_{0} + \frac{\mu}{s_{i}} + \epsilon_{i} , \quad \epsilon_{i} \sim N(0,1).
\label{eq:eggerPB}
\end{equation}

Testing for small study bias is then equivalent to testing $H_{0}$: $\beta_{0}$ = 0. If model (\ref{eq:eggerPB}) and $\ci_{(s_{i},\epsilon_{i})}$ hold, then the overall effect estimate, $\hat{\mu}$, adjusted for possible small study bias (via $\hat{\beta}_{0}$) is a consistent estimate for the overall treatment effect, $\mu$.  Several authors have considered the addition of random effects into model (\ref{eq:eggerPB}), in order account for possible residual heterogeneity after adjustment for small study bias, see for example and Moreno et al, (2009), Peters et al, (2010) and  R$\ddot{u}$cker et al, (2011). Their approaches have been straightforward generalisations of the additive and multiplicative random effects models (\ref{eq:RE}) and (\ref{eq:MRE}) respectively, as below:

\begin{eqnarray}
\frac{y_{i}}{s_{i}} &=& \beta_{0} + \frac{\mu}{s_{i}} + \frac{\delta_{i}}{s_{i}} + \epsilon_{i} ,  \quad \delta_{i} \sim N(\mu,\tau^{2}) \quad \epsilon_{i} \sim N(0,1). \label{eq:eggerPBRE} \\
\frac{y_{i}}{s_{i}} &=& \beta_{0} + \frac{\mu}{s_{i}} + \phi^{\frac{1}{2}}\epsilon_{i} ,  \quad \epsilon_{i} \sim N(0,1). \label{eq:eggerPBMRE}
\end{eqnarray}

Again, regardless of whether model (\ref{eq:eggerPBRE}) or (\ref{eq:eggerPBMRE}) holds in practice, under $\ci_{(s_{i},\delta_{i},\epsilon_{i})}$, consistent estimation of the overall mean parameter $\mu$ follows from fitting the standard fixed effect Egger model (\ref{eq:eggerPB}).\\
\\
At first sight,  model (\ref{eq:eggerPBRE}) seems the most natural extension to model (\ref{eq:eggerPB}). However, when a non-zero value for $\tau^{2}$ is estimated under model (\ref{eq:eggerPBRE}), the resulting overall estimate for $\mu$ differs from, and can often exhibit more substantial bias than, the fixed effect estimate (R$\ddot{u}$cker et al. (2011)). By contrast, multiplicative model (\ref{eq:eggerPBMRE}) is far more well behaved in this respect since its point estimate is identical to that obtained from fitting model (\ref{eq:eggerPB}). Nullifying the influence of variance components on the overall mean, a property enjoyed by model (\ref{eq:eggerPBMRE}), is so attractive in the presence of small study bias, that approaches have been developed to artificially incorporate this feature into additive random effects models as well (Henmi and Copas, 2010).\\
\\
For the reasons outlined above, we analyse the Aspirin data using the multiplicative Egger regression model only. This is straightforward, because model (\ref{eq:eggerPBMRE}) is {\it automatically} fitted by the process of standard linear regression using the least squares criterion. For these data  $\hat{\beta}_{0}$ = 2.11, with a p-value of approximately $1\times10^{-8}$, $\hat{\mu}$ is equal to 0.025, with a p-value of 0.89 and $\hat{\phi}$ = 0.64. In summary, Egger regression detects a highly significant presence of small study bias and, after this has been removed, no evidence of a treatment effect whatsoever. A slightly concerning feature is the severe under-dispersion in the residual error after adjustment for small study bias, an issue we return to in Section 5.

\section{Mendelian randomization and meta-analysis}

\begin{figure}[hbtp]
\begin{center}
\includegraphics[bb = 19 150 700 400,width=0.7\textwidth,clip]{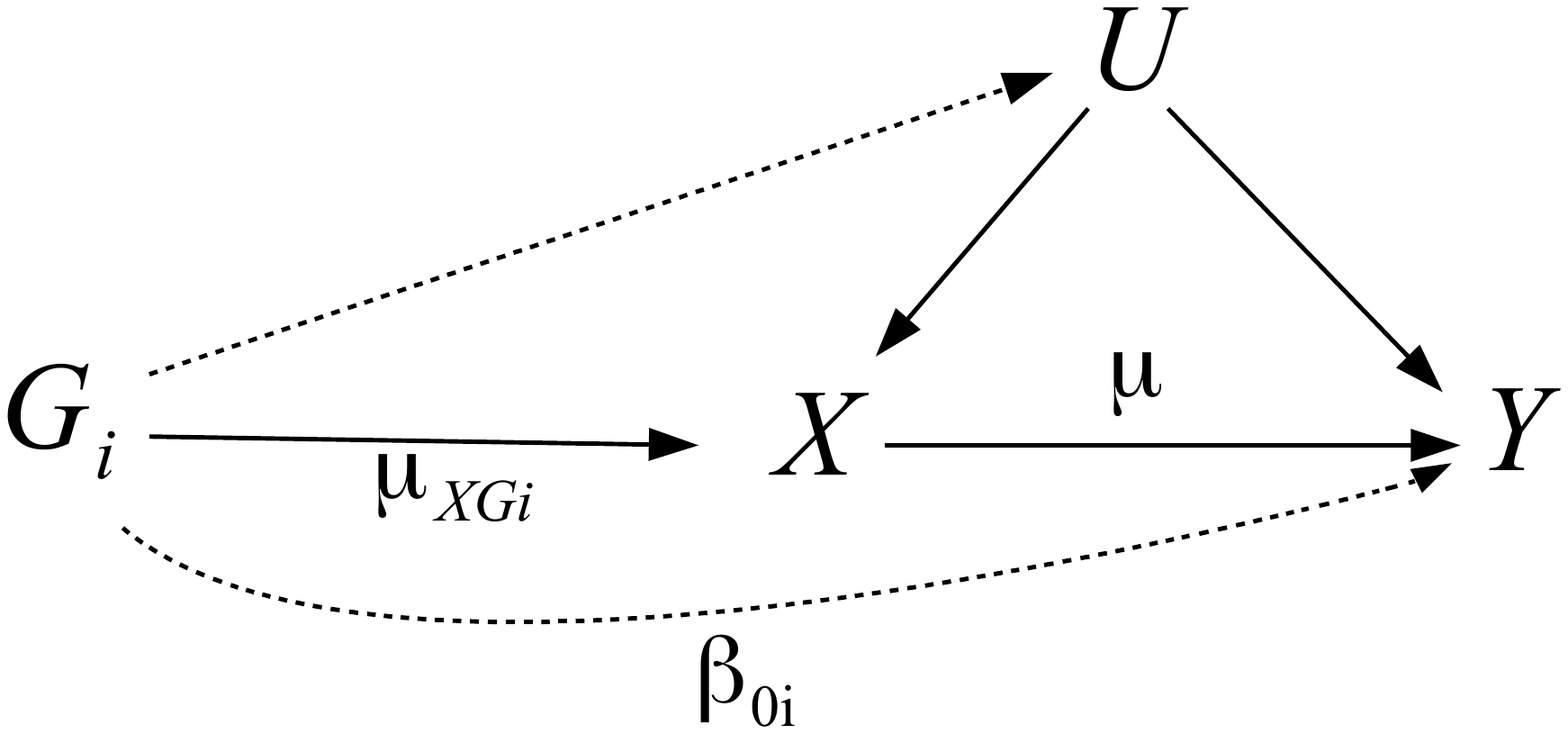}
\captionof{figure}{{\it Causal diagram representing the standard IV assumptions (solid lines), versus violations of the assumptions (dotted lines) }}
\label{fig:CausalDag}
\end{center}
\end{figure}

Mendelian randomization (MR) (Davey Smith and Ibrahim, 2003) is a technique applied to observational data, that exploits information on genetic variants to infer whether a modifiable risk factor, $X$, has a causal effect on a health outcome, $Y$. Straightforward analysis methods with observational data are vulnerable to confounding bias, and estimates of association can only have a true causal interpretation if all confounders between $X$ and $Y$ (which we denote by $U$) have been adjusted for. An MR study can also be used in place of a randomized clinical trial, when such a trial can not be implemented for ethical, practical or financial reasons. The method requires that a genetic variant, $G_{i}$, exists that satisfies the `Instrumental Variable' (IV) assumptions, see Bowden and Turkington (1990) for a thorough review. Specifically, $G_{i}$ must be: (i) associated with $X$; (ii) not associated $U$; and (iii) not associated with $Y$ given $X$ and $U$. These assumptions are encoded by the solid arrows in Figure \ref{fig:CausalDag}. The causal effect of $X$ on $Y$ is the parameter of interest and, following the notation in this paper, is denoted by $\mu$.
The Wald estimator is the ratio of the gene $i$ outcome association, $\hat{\mu}_{YGi}$, and the gene $i$ exposure association, $\hat{\mu}_{XGi}$, giving $\hat{\mu}_{i} = \frac{\hat{\mu}_{YGi}}{\hat{\mu}_{XGi}}$. Under the IV assumptions, $\hat{\mu}_{YGi}$ tends towards the product of the gene-exposure association and the causal effect, $\mu\hat{\mu}_{XGi}$, as the sample size grows large so that the Wald estimate it is asymptotically unbiased.\\
\\
Estimation of $\mu$ via a MR analysis incorporating multiple uncorrelated genetic variants $G_{1},...,G_{k}$, can be viewed as equivalent to a fixed effect meta-analysis of the Wald ratio estimates $\hat{\mu}_{1},...,\hat{\mu}_{k}$ (Johnson, 2014). So, formula (\ref{eq:FEest}) can be applied in the MR context to yield
\begin{equation}
\hat{\mu} = \frac{\sum^{k}_{i=1}\hat{\mu}^{2}_{XGi}\hat{\mu}_{i}}{\sum^{k}_{i=1}\hat{\mu}^{2}_{XGi}}
\label{eq:IVW}
\end{equation}
Relating this back to the general meta-analysis context, we can equate the gene-exposure association, $\hat{\mu}_{XGi}$, to a within study precision, $1/s_{i}$, and the causal effect estimate, $\hat{\mu}_{i}$, with the effect estimate for study $i$, $y_{i}$. Just as for a fixed effects meta-analysis, formula (\ref{eq:IVW}) assumes that $\hat{\mu}_{XGi}$, although estimated, is known. It also assumes for simplicity that all variants have the same allele frequency, but this can easily be accounted for if false. Formula (\ref{eq:IVW}) has the advantage of being calculable directly from summary data estimates, not therefore requiring data at the individual level.

\subsection{Pleiotropy and small study bias}

In a short period, the number of genetic variants with summary data estimates available for a given MR study has gone from a handful to, in many cases, several hundred, thus fuelling an exponential growth in the technique's application and its statistical power. However, as the number of genetic variants utilized in a MR analysis increases, so too does the likelihood that some of the included variants are invalid instrumental variables, which could potentially bias the analysis. The dotted lines in Figure \ref{fig:CausalDag} highlight these possible violations, namely (ii) that a variant may be associated with a confounder of $X$ and $Y$ and (iii), that a variant may affect the outcome via a completely independent pathway than $X$. Violation of assumptions (ii)  and (iii) can not be directly tested, and together the magnitude of their violation is referred to as the {\it pleiotropic} effect of a variant. Substituting $1/s_{i}$ for $\hat{\mu}_{XGi}$, $y_{i}$ for $\hat{\mu}_{i}$ and denoting the total pleiotropic effect of variant $i$ by $\beta_{0i}$, we can write the following more plausible model for the $i$'th Wald ratio estimate in a MR analysis using the notation of this paper as follows:
\begin{eqnarray}
\frac{y_{i}}{s_{i}} &=& \beta_{0i} + \frac{\mu}{s_{i}} + \epsilon_{i}.  \nonumber \\
                    &=& \beta_{0} + \frac{\mu}{s_{i}} +  \epsilon_{i} + \psi_{i}, \quad \epsilon_{i} \sim N(0,1) \quad \psi_{i} \sim N(0,\sigma^{2}_{\beta_{0}}) \label{eq:MREgger}
\end{eqnarray}
Model (\ref{eq:MREgger}), first conceptualized by Bowden, Davey Smith and Burgess (2015) but clarified here, allows the causal effect estimate of variant $i$ to be composed of a common causal effect, $\mu$, and an additional, individual pleiotropy term, $\beta_{0i} = \beta_{0}+\psi_{i}$. For convenience we assume that $\psi_{i}$ is normally distributed with zero mean and variance $\sigma^{2}_{\beta_{0}}$. The standard method of analysis for estimating the overall causal effect -- formula (\ref{eq:IVW}) -- assumes all genetic variants are valid IVs ($\beta_{0i}$ = 0 for all $i$). Egger regression implemented via model (\ref{eq:eggerPB}) -- which ignores the true heterogeneous nature of the bias by estimating only its mean value, $\beta_{0}$ -- can still consistently estimate the causal effect, $\mu$, even when {\it all} genetic variants exhibit pleiotropy, as long as $\ci_{(\psi_{i},s_{i},\epsilon_{i})}$ holds for model (\ref{eq:MREgger}). Under this assumption:
\begin{equation}
\hat{\mu} = \frac{\text{Cov}(y,\frac{1}{s})}{\text{Var(1/s)}} = \mu + \frac{\text{Cov}(\epsilon,\frac{1}{s})}{\text{Var(1/s)}} + \frac{\text{Cov}(\psi,\frac{1}{s})}{\text{Var(1/s)}} \label{eq:BiasFormula},
\end{equation}
and as the number of variants $k$ grows large, the covariance terms above tend to 0 and $\hat{\mu} \rightarrow \mu$. Bowden, Davey Smith and Burgess (2015) show that $\ci_{(\psi_{i},s_{i},\epsilon_{i})}$ is only plausible if IV assumptions (i) and (ii) hold. If assumption (ii) is violated (as indicated by the dotted line from $G_{i}$ to the confounder $U$ in Figure \ref{fig:CausalDag}),  $\psi_{i}$ and $s_{i} = 1/\gamma_{i}$ will contain common terms via re-introduced confounding and will be positively correlated, thus Egger regression will no longer provide a consistent estimate for $\mu$.

\section{A causal interpretation of Egger regression}

The application of Egger regression to a causal inference problem like Mendelian randomization is surprising, since this field has traditionally been dominated by instrumental variable methods that are conceptually very different. One such technique is the Structural Mean Model framework, which uses potential outcomes (Rubin, 2005) and places independence at the heart of its estimation strategy via the technique of G-estimation. For a general overview of this method see Vansteelandt and Joffe (2014) and, within the context of Mendelian randomization, Bowden and Vansteelandt, (2011). By making an analogy with Structural Mean Models, we now show that Egger regression can be understood as a means to de-bias a meta-analysis by restoring symmetry to the funnel plot, in a different but complimentary way to Trim and Fill. We start by assuming equation (\ref{eq:eggerPBMRE}) as a `working' mean model, even if in reality we believe that the heterogeneous bias model (\ref{eq:MREgger}) is more realistic. We multiply each side of the working model by $s_{i}$ and subtract $\beta_{0}s_{i}$ to yield
\[\
y_{i}(\beta_{0}) = y_{i} - \beta_{0}s_{i} = \mu + \phi^{\frac{1}{2}}\epsilon_{i}s_{i}.
\]
The term $y_{i}(\beta_{0})$ is a transformed version of the effect size estimates, can be viewed as a potential outcome, and is theoretically mean independent of $s_{i}$ under $\ci_{(s_{i},\epsilon_{i})}$.  The intercept estimate obtained from fitting model (\ref{eq:eggerPBMRE}), $\hat{\beta}_{0}$,  can be viewed as defining a particular transform of the data $y_{i}(\hat{\beta}_{0})$ that forces $y_{i}(\hat{\beta}_{0})$ to be independent of $s_{i}$ across all studies. Taking this further, model (\ref{eq:eggerPBMRE}) is equivalent to solving the following system of estimating equations:

\begin{eqnarray}
\text{Weight equation:} \quad w_{i} &=& 1/(\phi s^{2}_{i}) \nonumber \\
\text{Potential outcome transform:} \quad y_{i}(\beta_{0}) &=& y_{i} - \beta_{0}s_{i} \hspace{1cm} \nonumber \\
\text{Mean equation:} \quad\sum^{k}_{i=1}w_{i}\left\{y_{i}(\beta_{0}) - \mu\right\} &=& 0 \label{eq:meanJt} \\
\text{G-estimation equation:} \quad \sum^{k}_{i=1}w_{i}\left\{y_{i}(\beta_{0}) -\mu\right\}\left(s_{i} - \overline{s}\right)&=& 0 \label{eq:Gest} \\
\text{Heterogeneity equation:} \quad \sum^{k}_{i=1}w_{i}\left\{y_{i}(\beta_{0})-\mu\right\}^{2} - (k-2) &=& 0,  \label{eq:QEggerRE}
\end{eqnarray}

where $\overline{s}$ is the arithmetic mean of the $s_{i}$ terms. We now clarify the connection between the above system of estimating equations and estimation of $\beta_{0}$, $\mu$ and $\phi$ using standard linear regression theory. Fitting model (\ref{eq:eggerPB}) to obtain estimates for $\beta_{0}$ and $\mu$, is equivalent to solving equations (\ref{eq:meanJt}) and (\ref{eq:Gest}) leaving $\phi$ unspecified (in the Appendix we provide some simple \verb R  code to verify this). We then formally define $\phi$ as a parameter and solve equation (\ref{eq:QEggerRE}) to give
\begin{equation}
\hat{\phi} = \frac{\sum^{k}_{i=1}w_{i}\left\{y_{i}(\hat{\beta_{0}})-\hat{\mu}\right\}^{2}}{k-2}. \label{eq:Qdash}
\end{equation}

We note the equivalence of the numerator of equation (\ref{eq:Qdash}) to the Q$^{'}$ statistic defined in R$\ddot{u}$cker et al. (2011). The variances of $\hat{\beta}_{0}$ and $\hat{\mu}$ are given by

\[\
\text{Var}(\hat{\mu}) = \frac{\hat{\phi}}{\sum^{k}_{i=1}(1/s_{i} - \overline{1/s})^2}, \quad \text{Var}(\hat{\beta}_{0}) = \text{Var}(\hat{\mu})\overline{s^{2}},
\]

where $\overline{s^{2}}$ is the sample mean of the squared, within study variances.  \\
\\
Concerns over the use of Egger regression have been raised when analysing binary data because a study's outcome estimate will not truly be independent of its standard error, even when small study bias is not present. For this reason, Peters et al (2010) propose to replace $s_{i}$ in the Egger regression equation with a measure of precision based on study size, $1/n_{i}$ say. This could easily be represented in our estimating equation with suitable modification. For example, the G-estimation formula above would then be used to find the $\beta_{0}$ that forces independence between $y_{i}(\beta_{0})$ = $y_{i} - \beta_{0}/n_{i}$ and $1/n_{i}$.

\subsection{Re-analysis of  the Aspirin data}

\begin{figure}[hbtp]
\begin{center}
\includegraphics[width=0.49\textwidth,clip]{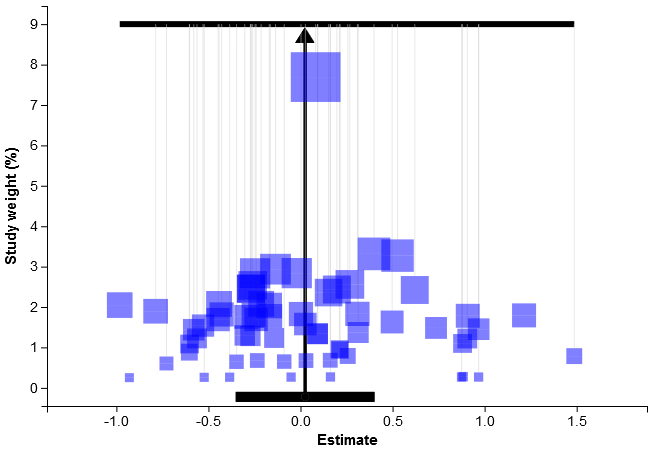}
\includegraphics[width=0.49\textwidth,clip]{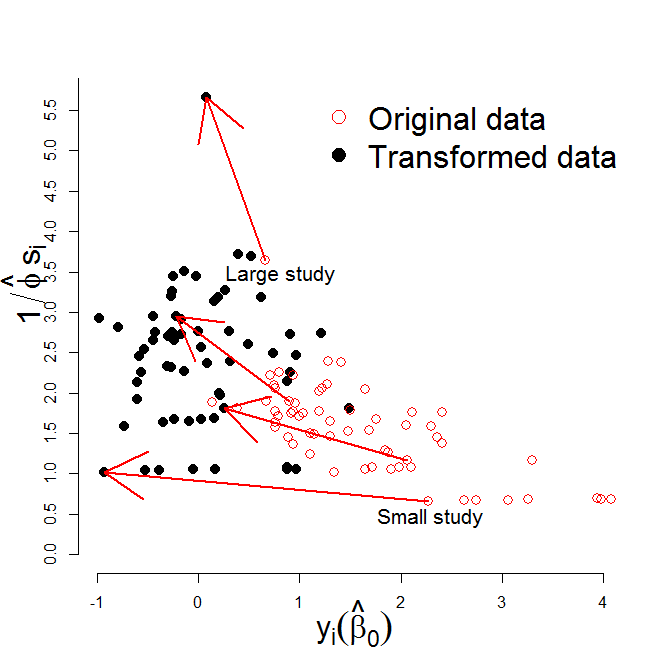}
\caption{{\it Left: The Meta-Analyzer incorporating Egger regression enacted on the Aspirin data and shown under the potential outcome transform. Right: Funnel plot of the original Aspirin data ($y_{i}$ vs $1/s_{i}$, hollow red dots) versus its transformed counterpart ($y_{i}(\hat{\beta}_{0})$ vs $1/\hat{\phi}s_{i}$, solid black dots). }}
\label{fig:ShinyFunnel3}
\end{center}
\end{figure}

Returning to the Aspirin data, Figure \ref{fig:ShinyFunnel3} (left) shows the final resting point of the Meta-Analyzer upon enacting Egger regression using the causal interpretation described above. Once the Egger regression option is ticked, users observe a dynamical change in the Meta-Analyzer from its initial starting point of the standard random effects analysis in Figure \ref{fig:ShinyFunnel2}. The x-axis position is now the transformed or potential outcome scale $y_{i}(\hat{\beta}_{0})$ = $y_{i}(2.11)$ for study $i$. The meta-analysis now exhibits a high amount of symmetry that can be immediately visualised by the user. This transformation is highlighted in Figure \ref{fig:ShinyFunnel3} (right), which plots potential outcomes $y_{i}(\hat{\beta}_{0})$ on the horizontal versus $1/\hat{\phi}s_{i}$ on the vertical axis with the original data ($y_{i}$ versus 1/$s_{i}$) also shown for comparison. When small study bias is present, estimates from small studies are shifted by a relatively large amount in the horizontal dimension (in this case to the left), whereas those from large studies are shifted horizontally by a relatively small amount.\\
\\
Because $\phi$ is estimated to be 0.64 in this instance, indicating under-dispersion after adjustment for small study bias, study precisions under the potential outcome transformation are increased in proportion to their size, so that the precisions of small studies are shifted vertically upwards by a small amount and those of large studies are shifted by a large amount. If over-dispersion had been observed after adjustment, we would have seen a shift vertically downwards instead. If we think the that heterogeneous bias model (\ref{eq:MREgger}) is true and $\ci_{(\psi_{i},s_{i},\epsilon_{i})}$ holds, then we would expect these data to be over-dispersed since, from equation (\ref{eq:Qdash}), $E[\hat{\phi}]$ = $1+\sigma^{2}_{\beta_{0}}$. Therefore, we can only meaningfully interpret $\hat{\phi}$ - 1 as an estimate for $\hat{\sigma}^{2}_{\beta_{0}}$ if  $\hat{\phi} > $ 1.

\subsubsection{Can Egger regression correct for publication bias?}

The heterogeneous bias model (\ref{eq:MREgger}) seems well suited to the MR context. Furthermore, the work of Hedges and Copas {\it do} indeed imply that heterogeneous bias contaminates the outcome data, $y_{i}$, under publication bias. This is because the selective pressure study $i$ experiences (and therefore the magnitude of bias in its estimate conditional on selection) is a function of its own, unique characteristics, such as $s_{i}$. So, whilst model (\ref{eq:MREgger}) does therefore appear to be a realistic data generating model for mainstream meta-analyses affected by small study bias, if model (\ref{eq:MREgger}) is true then $\ci_{(\psi_{i},s_{i},\epsilon_{i})}$ will be implausible because $\psi_{i}$ will clearly then be a correlated with $s_{i}$. Ironically, Egger regression is therefore unlikely to adequately adjust for small study bias when induced by outcome dependent sampling, such as in the clinical trial context (for which it was originally intended). Since under-dispersion is a phenomenon that can also easily be induced by publication bias (Copas and Shi, 2000) this provides even stronger evidence that the Egger regression analysis of the Aspirin data should be interpreted with extreme caution.

\subsection{The causal effect of LDL cholesterol on heart disease risk}

\begin{figure}[hbtp]
 \centering
\includegraphics[width=0.49\textwidth,clip]{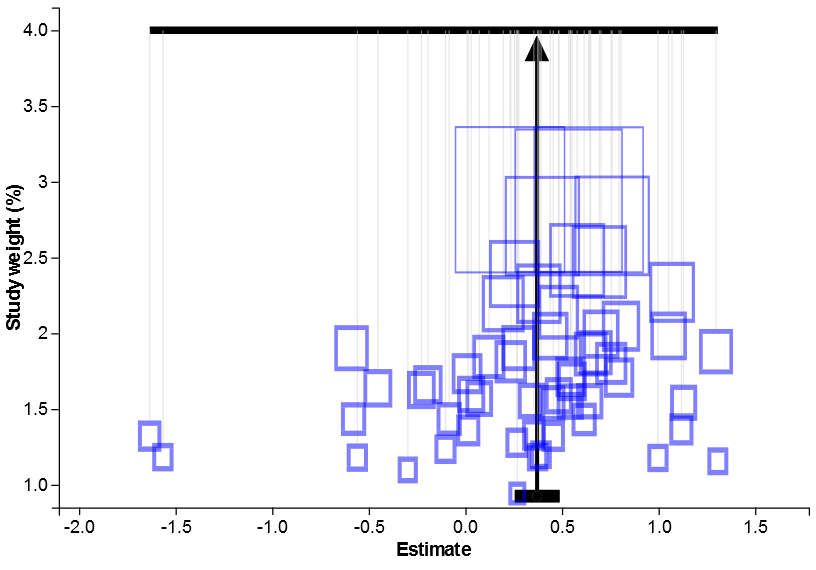}
\includegraphics[width=0.49\textwidth,clip]{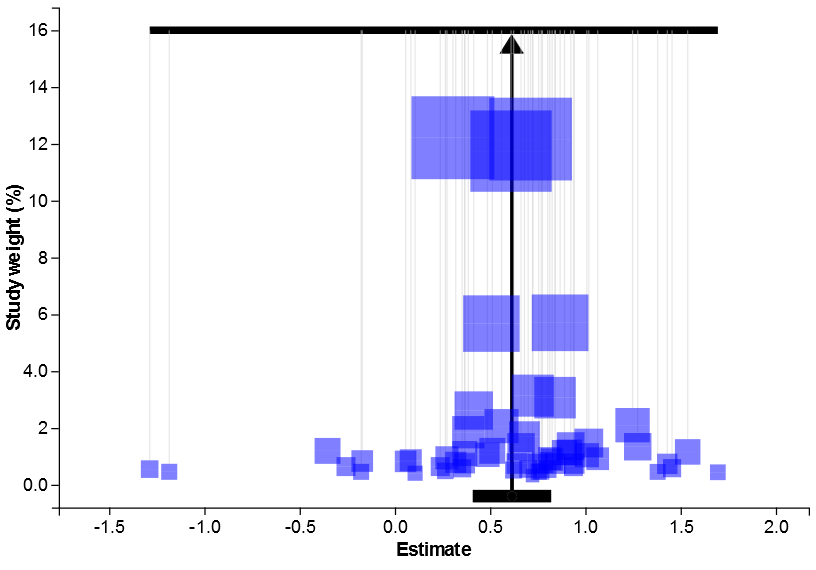}
\caption{{\it The Meta-Analyzer supporting an MR analysis of the LDL variants under an additive random effects model (left) and Egger regression (right).}}
\label{fig:LDLanalysis}
\end{figure}

There is a long an extensive literature on the association between various lipid fractions and cardiovascular disease, but still far from universal agreement as to whether these associations have a causal basis. Using summary data estimates available on 154 variants from the CARDIOGRAM consortium (CARDIOGRAM, 2013), we perform a Mendelian randomization analysis to look for evidence that low density lipoprotein cholesterol (LDL-c) has a causal effect on coronary artery disease (CAD).\\
\\
Since LDL-c levels are closely related and highly correlated with other lipid fractions, such as Triglycerides and high density lipoprotein, we selected only a subset of 57 variants out of the 154 that were most strongly associated LDL-c. The minimum p-value for the strength of association across all variants was 8.3e-7. This strategy should reduce the possibility of violating causal assumption (ii), but does not rule it out completely. Furthermore, the selected variants might well exhibit pleiotropic effects through completely separate pathways, and therefore be in violation of assumption (iii). For this reason, we supplement the standard IVW analysis with Egger regression under model (\ref{eq:MREgger}).\\
\\
Figure \ref{fig:LDLanalysis} (left) and Table \ref{tab:LdLResults} shows the result of a standard additive random effects meta-analysis applied to the causal effect estimates across all 57 included variants. They quantify the causal effect in terms of a log-odds ratio of coronary artery disease for a 1 standard deviation increase in LDL-c levels. Significant heterogeneity is detected in the data ($I^{2}$ = 70\%, $\tau^{2}$ = 0.11), despite this, a strong positive causal log-odds ratio of 0.37 is estimated. However, there is reason to believe this analysis to be misleading, given that there exists a noticeable correlation between the magnitude of the causal effect estimate and its precision, indicative pleiotropy. Removal of the two relatively imprecise causal effect estimates which are less than -1.5 reduces the heterogeneity considerably (adjusted $\tau^{2}$=0.07, results not shown).

\begin{table}[htbp]
\begin{center}
\begin{tabular}{lcccc}
\hline
Model & \\
Parameter & Est & S.E & t value & p-value \\
\hline
\multicolumn{5}{c}{Random effects model (3)} \\
$\mu$  & 0.37 &   0.059 &  6.29 & 5.10e-08 \\
$\tau^{2}$ & 0.11 &-&-&-\\
\multicolumn{5}{c}{Random effects model (4)} \\
$\mu$  & 0.45 &   0.053 &  8.51 & 1.13e-11 \\
$\phi$ & 3.33 &-&-&-\\
\multicolumn{5}{c}{Egger regression model (12)} \\
$\beta_{0}$ &-0.0102&0.0046&-2.23&0.0298 \\
$\mu$       &0.632&0.0975&6.481&2.66e-08 \\
$\sigma^{2}_{\beta_{0}}$& 2.11 &-&-&-\\
\hline
\end{tabular}
\end{center}
\caption{\label{tab:LdLResults} {\it MR-Analyses of the lipids data. Estimates for $\mu$ are log-odds ratios per standard deviation increase in LDL-c.}}
\end{table}

Applying the multiplicative random effects model (4) to these data to remove the influence of the random effects variance, $\tau^{2}$, on the overall mean estimate, yields a slightly higher causal effect estimate of 0.45. Applying Egger regression model (\ref{eq:MREgger}) -- Figure \ref{fig:LDLanalysis} (right) and Table \ref{tab:LdLResults} -- a significant negative effect of pleiotropy is detected, despite the pleiotropy variance being large ($\hat{\sigma}^{2}_{\beta_{0}}$ = 2.11). Consequently, the point estimate for $\mu$ is adjusted upward to 0.63. In conclusion, although significant evidence of pleiotropy exists across the included variants, there is still overwhelming evidence that LDL-c is causally related to CAD risk. If anything, the Egger analysis suggests that the true causal effect of LDL-c is slightly masked by pleiotropy acting in the opposing direction.

\section{Discussion}

In this paper we have shown that, by augmenting the funnel plot portraying a meta-analysis of study results with an additional pole, cord and pivot, we can give this abstract object -- and the overall estimate that it implies -- a clear physical interpretation. From its original conception and success as a science festival exhibit, we think the Meta-Analyzer has the continued potential to be a useful tool for educating and explaining the concept of weighing evidence and statistical reasoning to an even wider lay--audience, and hopefully to inspire the next generation of data scientists.\\
\\
We have shown that the conceptual framework the Meta-Analyzer promotes is useful for both explaining the rationale, and interpreting the effect of, extended modelling choices in meta-analysis to a more technical, advanced audience. Finally, it has allowed us to make connections between methods to adjust for small study bias in meta-analysis, and for confounding bias in  observational epidemiology.\\
\\
It is worth commenting on the differing way in which heterogeneity is modelled in standard meta-analysis and causal inference. In the former, significant differences in effect estimates across studies (as measured by Cochran's $Q$ statistic, say) is seen to provide clear evidence of underlying treatment effect heterogeneity and the subsequent adoption of a random effects model. However in causal inference it is strongly assumed that all instrumental variables identify a common causal effect. Any evidence of heterogeneity between causal effect estimates across different genetic instruments in an MR analysis is seen as evidence of pleiotropy (or invalid instruments more generally), and formal testing procedures such as the Sargan test (Sargan, 1951) exist to identify the invalid variants responsible so that they can be removed from the analysis. In our investigation of the magnesium data we did indeed remove an an outlying study in order to demonstrate a dramatic reduction in the between trial heterogeneity. This is not encouraged within the general context of meta-analysis but perhaps, as in the causal inference field, it should be tolerated to a higher degree.\\
\\
As the subject of Mendelian randomization moves forward - facilitated by the emergence of increasingly rich summary data sources, it is vital that the field makes use of existing methodology developed for meta-analysis, particularly in the area of bias modelling. However, by viewing the assumptions of established methods like Egger regression from a causal perspective, it is also possible that new insights can be gleaned and related back to the field of meta-analysis more generally. We hope that our paper can act as a catalyst to help promote and further this aim.

\section*{Acknowledgements}
We Thank the MRC visual aids department for their help in building the Meta-Analyzer, the MRC Biostatistics Unit staff members who demonstrated its use at the Science fair and to Vikki O'neill in particular. We also thank Philip Haycock for help in accessing the CARDIoGRAM data, as well as Dan Jackson and George Davey Smith for their helpful comments on an early draft of this work. Dr Jack Bowden is supported by an MRC methodology research fellowship (grant number MR/L012286/1).

\newpage

\appendix

\section*{The Meta-Analyzer from theory to practice}

\begin{framed}
\textbf{Box 1: Sherlock Holmes and the case of the missing evidence}%
\begin{quote}
`It is 1858. A new machine has been invented to exploit the ethereal force of gravity to weigh evidence: The Meta-Analyzer.  Its inventor, Stata Lovelace says {\it "The Meta-Analyser can make decisions affecting Britain and her empire in the name of gold-standard fairness and morality"}. The night before its maiden calculation (into the affect of sanitation on patient mortality: is it beneficial?), Professor Moriarty has stolen some evidence! Call Scotland Yard! And Sherlock Holmes!'\\
`Inspector Lestrade inspects the lop-sided Meta-Analyzer (Figure \ref{fig:MetaAMissing} (left)) and mops his brow'\\
{\bf Inspector Lestrade}: `It's simple 'olmes, just move the pivot point' (Figure \ref{fig:MetaAMissing} (middle)), `Case closed.'\\
{\bf Sherlock Holmes}: `Its a solution, but an inelegant one, Lestrade. Replace the missing studies to return the balance and yield the unbiased truth' (Figure \ref{fig:MetaAMissing} (right))
\end{quote}
\begin{center}
\includegraphics[bb = 19 300 600 812,width=0.3\textwidth,clip]{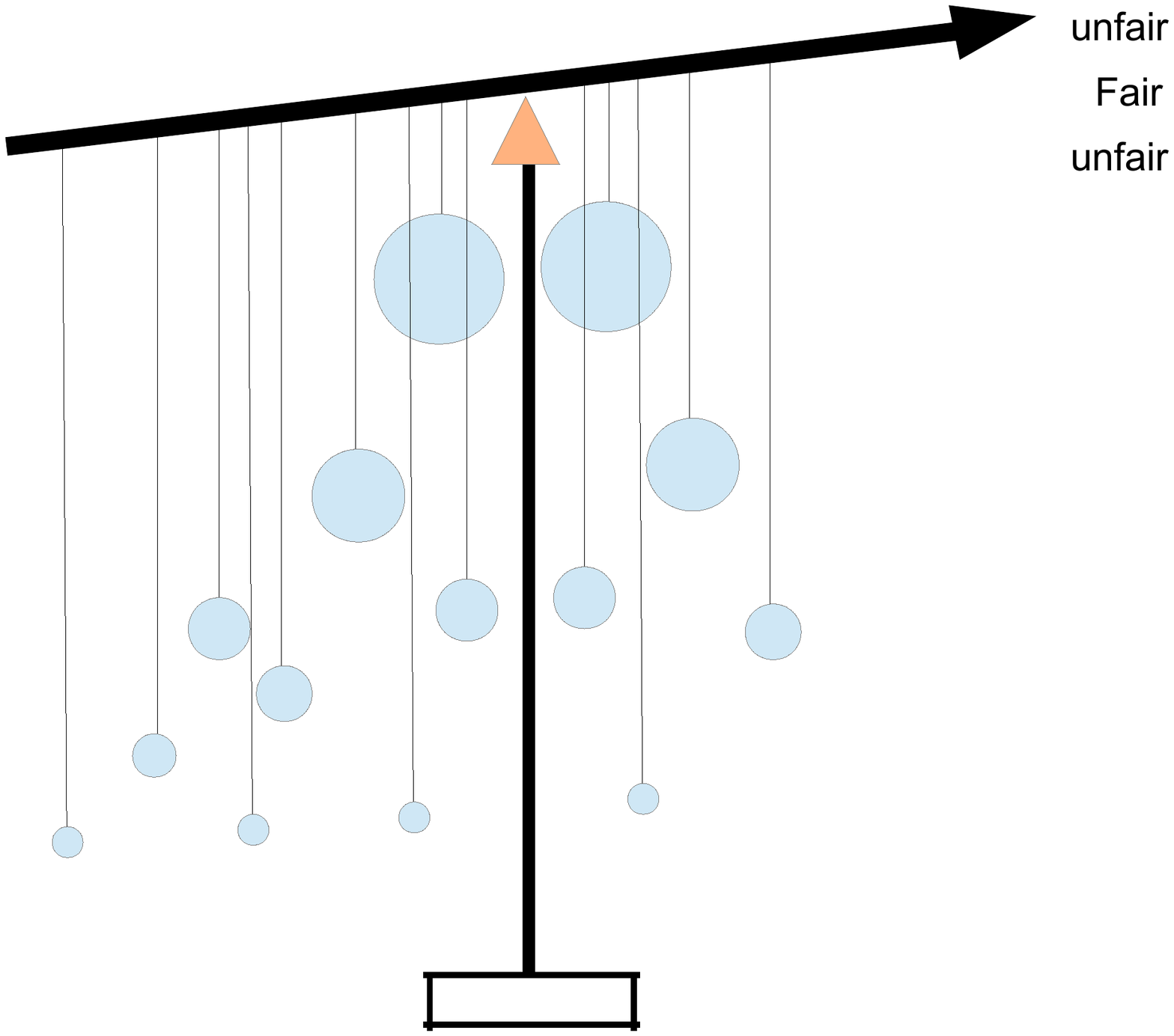}
\includegraphics[bb = 19 300 600 812,width=0.3\textwidth,clip]{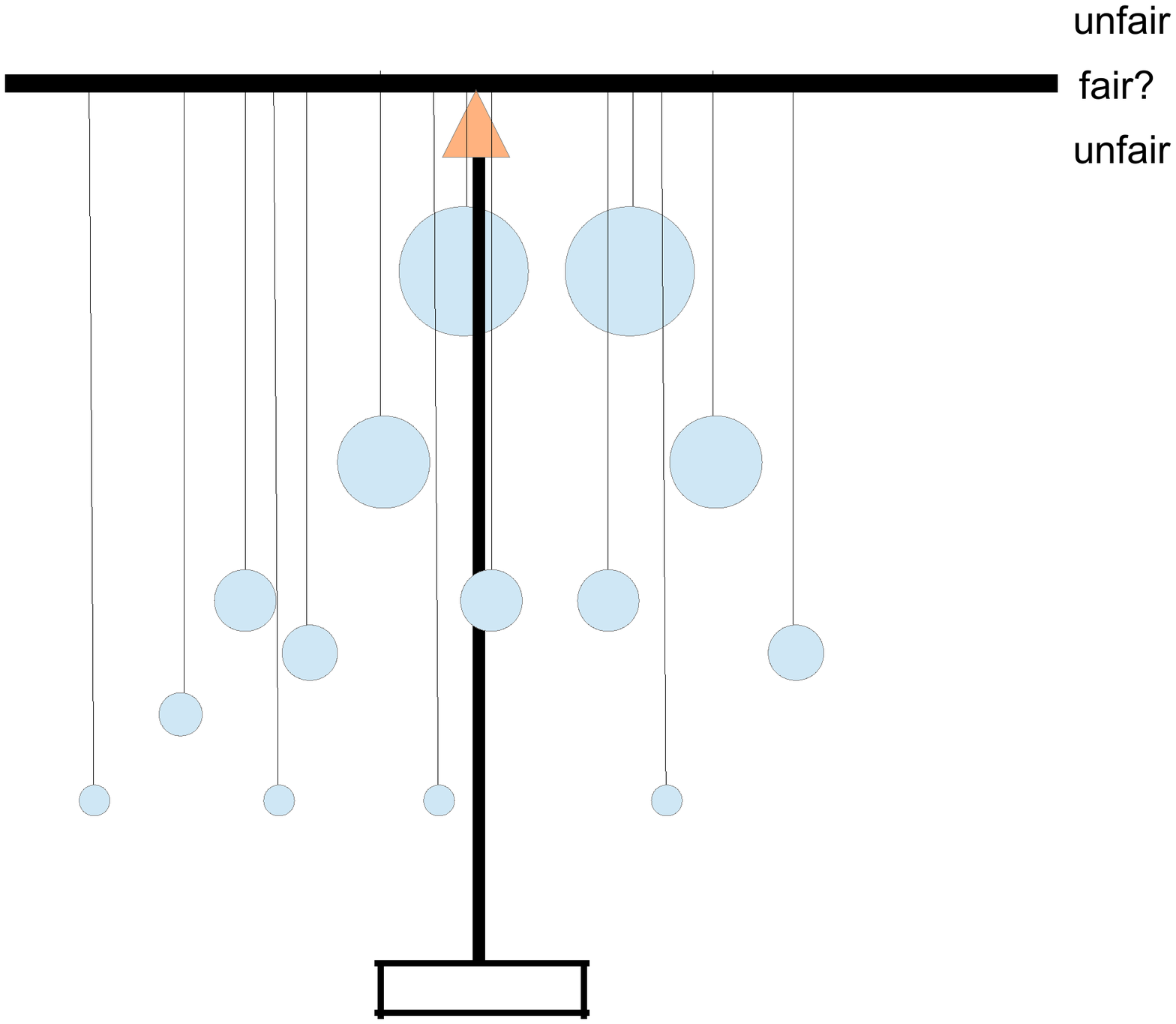}
\includegraphics[bb = 19 300 600 812,width=0.3\textwidth,clip]{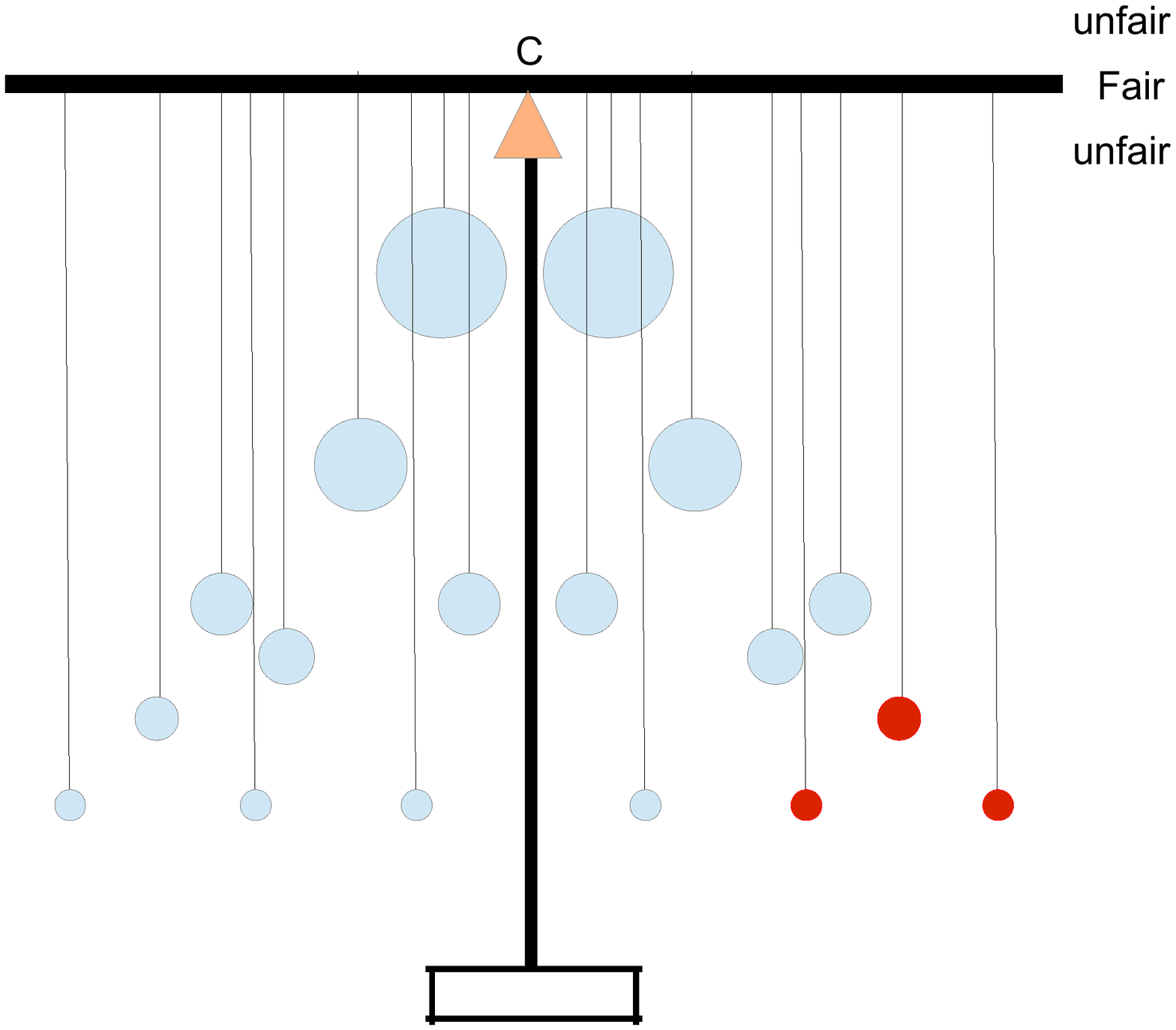}
\captionof{figure}{{\it Left: Unbalanced Meta-Analyzer. Middle: Rebalanced Meta-Analyzer (Inspector Lestrade approach). Right: Rebalanced Meta-Analyzer (Sherlock Holmes approach)}}
\label{fig:MetaAMissing}
\end{center}
\end{framed}

\newpage

\begin{figure}[hbtp]
 \centering
\includegraphics[width=0.6\textwidth,clip]{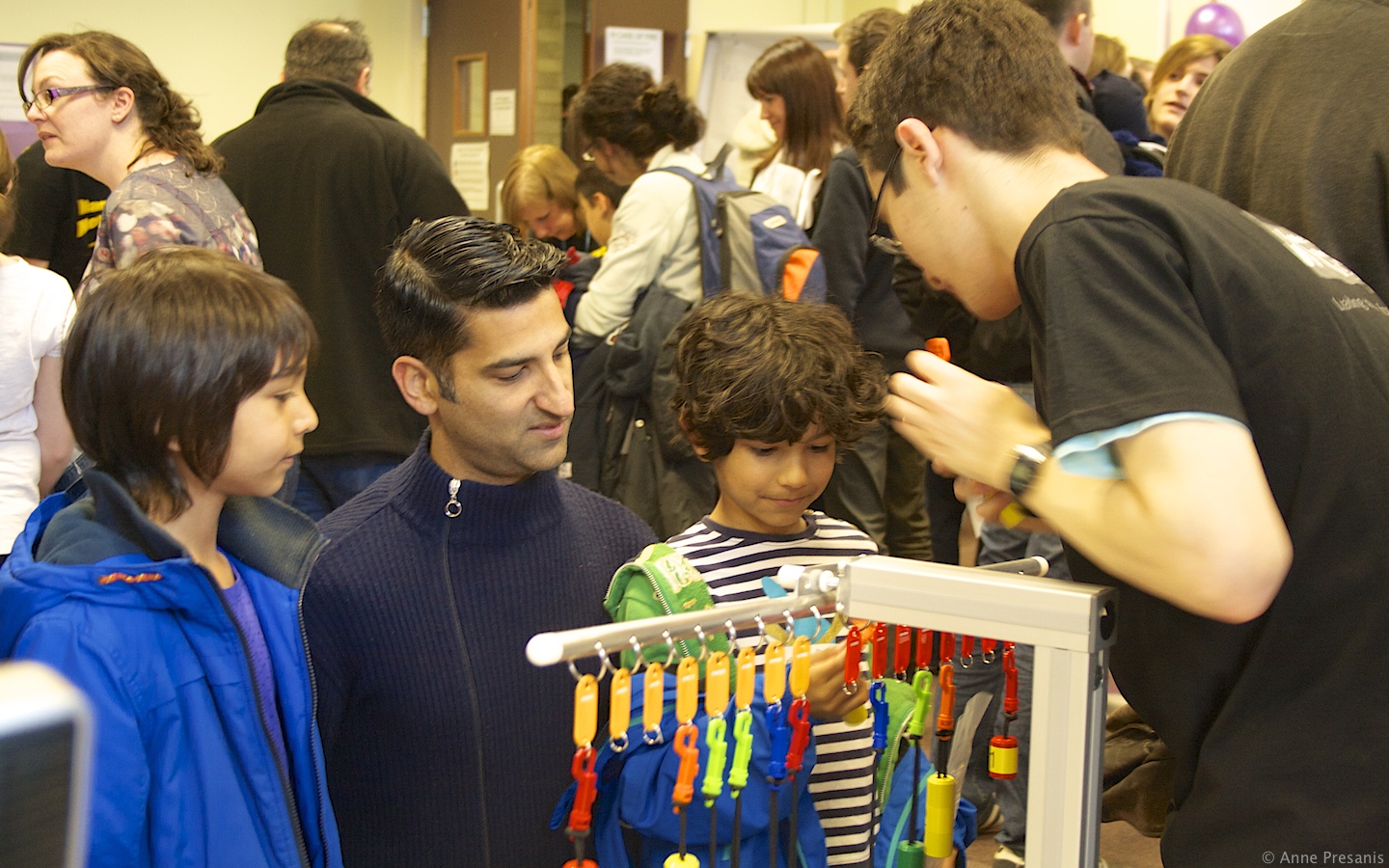}
\includegraphics[width=0.6\textwidth,clip]{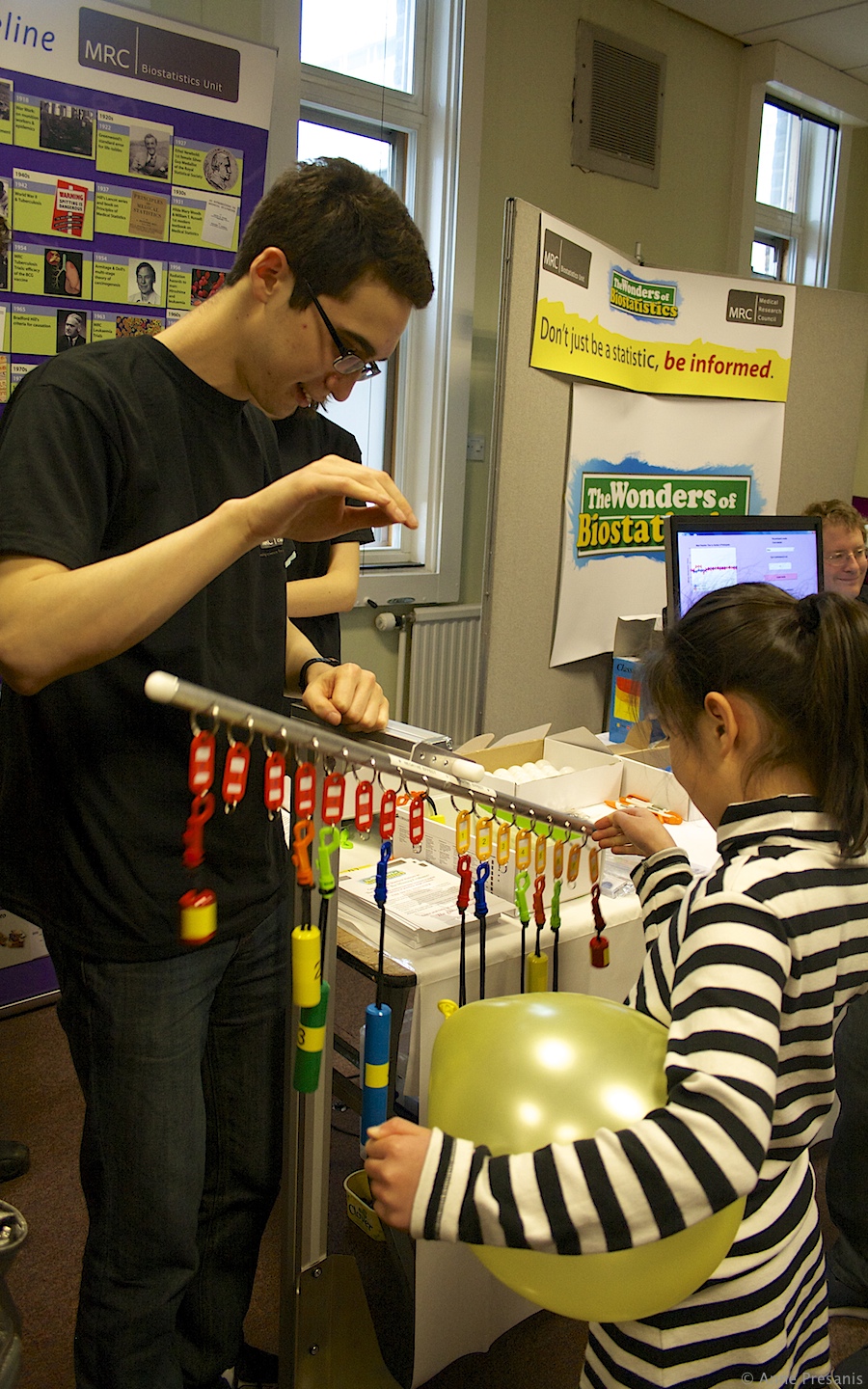}
\caption{{\it The Meta-Analyzer in action at the Cambridge Science Festival}}
\end{figure}

\newpage

\section*{R code for Section 5}

\begin{verbatim}
## given Aspirin data vectors y,s
## G-estimation routine
G_est = function(a){
w          = 1/s^2
Beta0      = a[1]
yBeta0     = y - Beta0*s
MU         = sum(w*yBeta0)/sum(w)
L          = (sum(w*(yBeta0-MU)*(s - mean(s))))^2
}
Beta0hat   = optimize(G_est,c(-5,5))$min
yBeta0hat  = y - Beta0hat*s
w          = 1/s^2
mu         = sum(w*yBeta0hat)/sum(w)
>Beta0hat
[1] 2.112803
> mu
[1] 0.02519816
> ## standard Egger regression
> summary(lm(y~s,weights=1/s^2))$coef[,1]
(Intercept)           s
 0.02519816  2.11280317
\end{verbatim}

\end{document}